\documentclass[superscriptaddress, amsmath, amssymb, aps,pra,nofootinbib,twocolumn
]{revtex4-2}
\usepackage{graphicx} 
\usepackage{comment}
\usepackage{dsfont}
\usepackage{color}
\usepackage{bm} 
\usepackage{hyperref}

\usepackage[utf8]{inputenc}
\usepackage{amsmath}
\usepackage{amssymb}
\usepackage[a4paper, total={6.6in, 8.8in}]{geometry}
\usepackage{float}
\usepackage{braket}
\usepackage{adjustbox}
\usepackage{color}
\usepackage{framed} 
\usepackage{url}
\usepackage[normalem]{ulem}

\usepackage[table]{xcolor}
\usepackage{colortbl}
\usepackage{geometry}
\definecolor{lightred}{RGB}{255, 200, 200}
\definecolor{lightgreen}{RGB}{200, 255, 200}

\begin{document}

\title{Tunable Random Telegraph Noise in Stable Perpendicular Magnetic Tunnel Junctions for Unconventional Computing
}

\author{Ahmed Sidi El Valli}
\affiliation{Center for Quantum Phenomena, Department of Physics, New York University, New York, NY 10003 USA}

\author{Michael Tsao}
\affiliation{Center for Quantum Phenomena, Department of Physics, New York University, New York, NY 10003 USA}

\author{Dairong Chen}
\affiliation{Center for Quantum Phenomena, Department of Physics, New York University, New York, NY 10003 USA}

\author{Andrew D. Kent}
\affiliation{Center for Quantum Phenomena, Department of Physics, New York University, New York, NY 10003 USA}

\begin{abstract}

We demonstrate that thermally stable perpendicular magnetic tunnel junctions (pMTJs), widely used in spin-transfer torque magnetic random-access memory, can be actuated with nanosecond pulses to exhibit tunable stochastic behavior. This actuated-stochastic tunnel junction (A-sMTJ) concept produces random telegraph noise, with control over fluctuation rate and probability bias. The device response is shown to be consistent with a Poisson process, with fluctuation rates tunable over more than two orders of magnitude, with average state dwell times varying from $29$ ns to greater than $2.3 \mu$s. These results establish A-sMTJs as a versatile platform for integrating deterministic, stochastic, and in-memory functionality on a single chip—advancing the development of probabilistic, neuromorphic, and unconventional computing systems.
\end{abstract}
\date{August 11, 2025}
\maketitle

\section{\label{sec:intro}Introduction}
The search for hardware building blocks that support unconventional computing architectures is a growing challenge at the interface of physics and information processing. Driven by rapid advances in artificial intelligence, there is increasing interest in addressing the memory bottlenecks of conventional von Neumann architectures by developing specialized hardware that exploits physical dynamics and brain-inspired computing models~\cite{Misra2023,Markovic2020,Nikhar2024,Finocchio2024}. One promising candidate is the perpendicular magnetic tunnel junction (pMTJ)---a device originally designed for fast, dense nonvolatile memory~\cite{Kent2015,Dieny2016}. More recently, pMTJs have been studied for in-memory computing~\cite{Jung2022,Sebastian2020,Hirtzlin2019}, and their intrinsic stochastic switching behavior has enabled applications in probabilistic computing and true random number generation (TRNG)~\cite{Rehm2023,Dubovskiy2024,Xu2024}.

In this work, we present a complementary approach by demonstrating that both non-volatile memory and superparamagnetic-like behavior can be achieved using pMTJs with stable magnetization free layers. Our results demonstrate that standard pMTJs can exhibit multifunctionality---serving not only as building blocks for conventional spin-transfer torque magnetic random access memory (STT-MRAM) in in-memory computing systems, but also as stochastic elements for probabilistic and unconventional computing architectures.

\begin{figure*}
 \centering
 \includegraphics[width=0.9\textwidth]{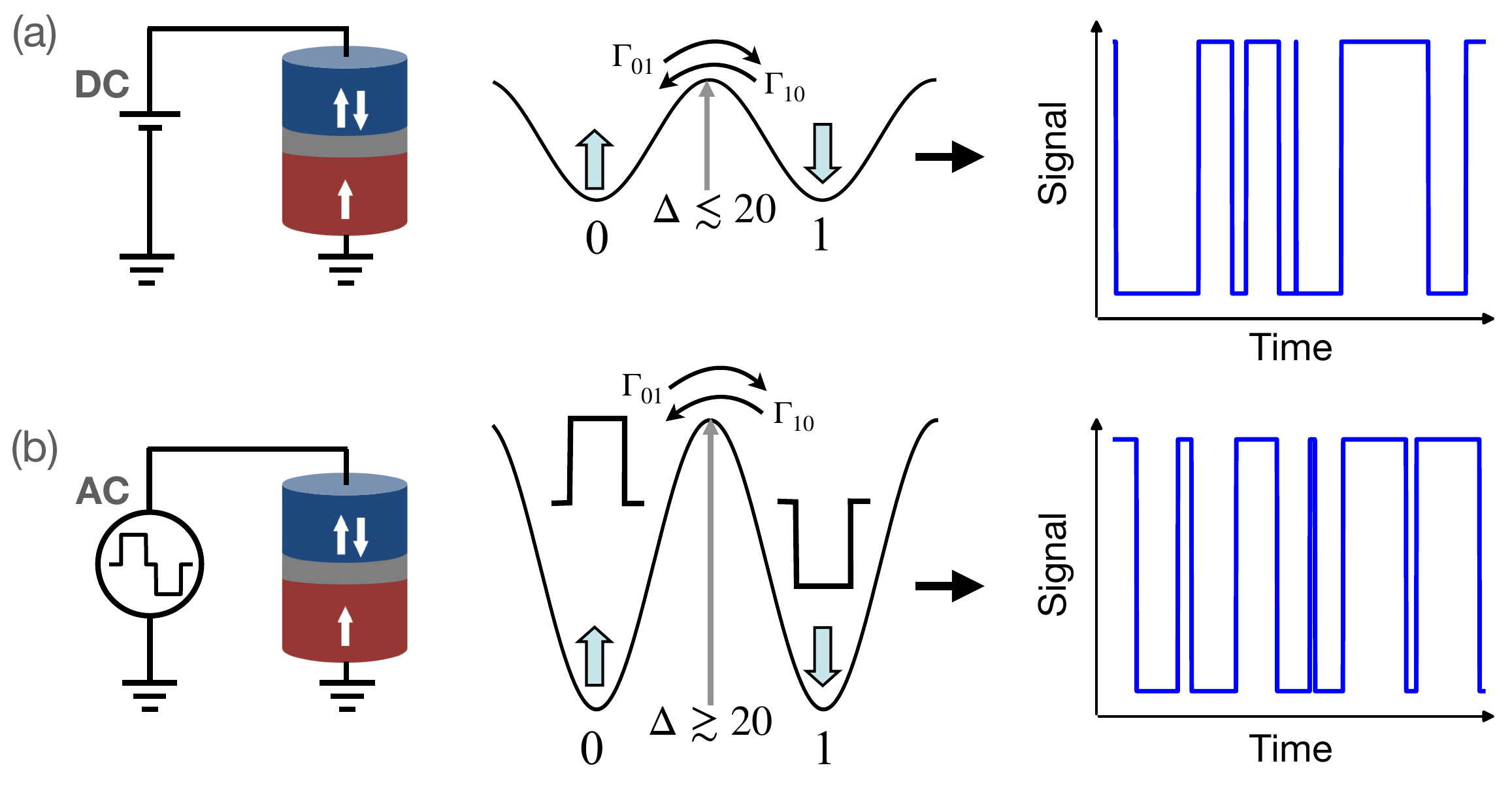}
 \caption{Stochastic magnetic tunnel junctions. (a) In superparamagnetic MTJs (sMTJs) with low energy barriers (\(\Delta \lesssim 20\)), thermal fluctuations spontaneously drive switching between the parallel (P, “0”) and antiparallel (AP, “1”) states, producing random telegraph signals in the device resistance. 
(b) In contrast, in actuated-stochastic magnetic MTJs (a-sMTJs), we suppress spontaneous switching by using pMTJs with larger energy barriers (\(\Delta \gtrsim 20\)) and actuate stochastic transitions via nanosecond-scale voltage pulses of alternating polarity. Here spin-transfer torques enable controlled, tunable switching, reproducing telegraph-like behavior through a fundamentally different, externally driven mechanism.}
\label{Fig:Schematic}
\end{figure*}

\section{Device Concept}
Stochastic pMTJ concepts are illustrated in Fig.~\ref{Fig:Schematic}. The key parameter governing the thermally activated switching behavior is the dimensionless ratio 
\(\Delta = E_B / (kT)\), 
where \(E_B\) is the energy barrier to magnetization reversal, \(k\) is the Boltzmann constant, and \(T\) is the device temperature. As shown in Fig.~\ref{Fig:Schematic}(a), when \(\Delta \lesssim 20\), the free layer magnetization undergoes spontaneous fluctuations between the parallel (P, labeled ``0'') and antiparallel (AP, labeled ``1'') states. The switching rate is governed at least approximately by an Arrhenius law, \(\Gamma \simeq \Gamma_0 \exp(-\Delta)\), where \(\Gamma_0\) is the attempt frequency, typically on the order of tens of GHz, but much higher rates have also been found in experiment~\cite{Bedau2010b,Soumah2025}. These thermal fluctuations generate random telegraph signals in the electrical resistance due to the tunnel magnetoresistance effect and are known as superparamagnetic magnetic tunnel junctions (sMTJs). 

In contrast, here we demonstrate an actuated MTJ device concept in which stochastic behavior is externally controlled. We use pMTJs with larger energy barriers (\(\Delta \gtrsim 20\)) where spontaneous thermally activated switching is suppressed.  In this regime, inspired by the device operation model presented in Ref.~\cite{Sun2024}, we induce stochastic switching by applying a periodic stream of nanosecond-duration voltage pulses of alternating polarity (hereafter referred to as ``AC pulses'' for brevity), as illustrated in Fig.~\ref{Fig:Schematic}(b). These pulses generate spin-transfer torques that modulate the magnetization dynamics, enabling fully controlled and tunable stochastic fluctuations. The switching probability depends on the amplitude and duration of the pulses and, under appropriate conditions, produces a random telegraph signal—functionally analogous to that of a superparamagnetic MTJ (sMTJ). We designate this as an actuated-stochastic magnetic tunnel junction (A-sMTJ), to distinguish it from the thermally driven fluctuations of the sMTJ shown in Fig.~\ref{Fig:Schematic}(a).

\section{Device Operation}
The device operating principle is based on the fact that spin-transfer torque switching is probabilistic, with the switching probability depending on the pulse polarity, amplitude and duration. The pulse polarity sets the switching sense, \emph{e.g.}, in the case of our experiment, positive pulse polarity favors P $\rightarrow$ AP (``0'' to ``1'') switching and negative pulse polarity AP $\rightarrow$ P (``1'' to ``0'') switching. For a given switching sense and collinear magnetization reference and free layers the switching probability is generally a monotonically increasing function of the pulse amplitude and duration (see, for example, Refs.~\cite{Liu2014,Rehm2019}). For STT-MRAM applications the pulse parameters are chosen to lead to highly reliable switching with very small write-error-rates (WER) ($<10^{-9}$ or lower depending on the application), termed a ``deterministic'' write pulse (see, for example, Ref.~\cite{Nowak2011}).  In our application the WER is chosen to be much larger to enable stochastic behavior. We denote these stochastic write pulses, to distinguish them from the write pulses used in STT-MRAM.

\begin{figure}
 \centering
 \includegraphics[width=0.4\textwidth]{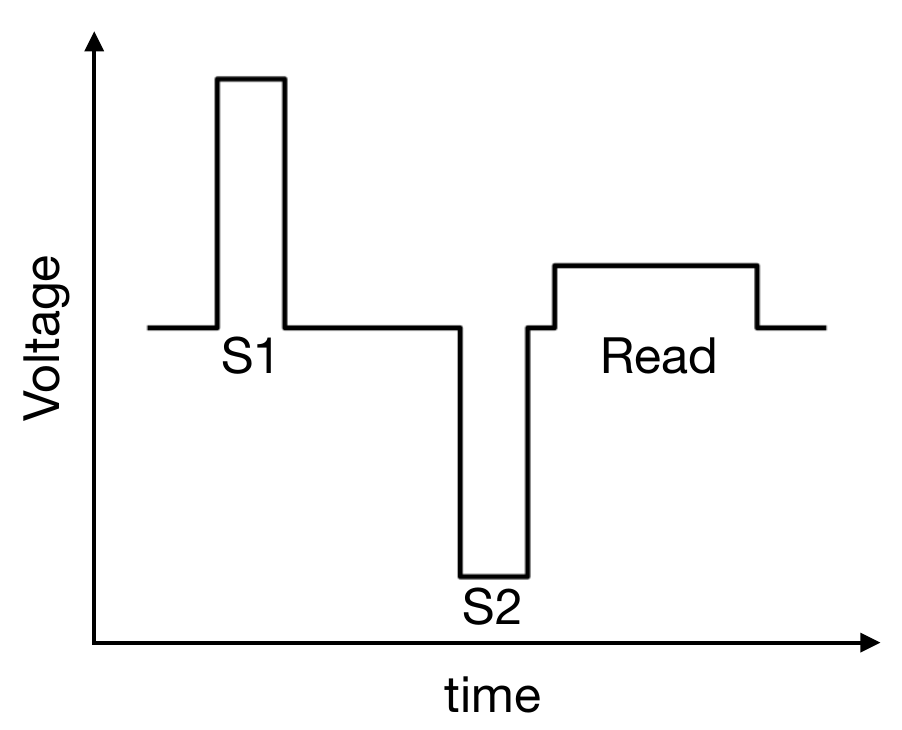}
 \caption{Voltage pulse sequence. Each cycle consists of a positive stochastic write pulse (S1) that promotes $0 \rightarrow 1$ transitions, followed by a negative stochastic write pulse (S2) that promotes $1 \rightarrow 0$ transitions. A subsequent low-amplitude read pulse is applied to detect the MTJ state.} 
 \label{Fig:pulse}
\end{figure}

Our device operates using two stochastic write pulses of opposite polarity, labeled S1 and S2, followed by a low-amplitude read pulse that does not perturb the MTJ state, as illustrated in Fig.~\ref{Fig:pulse}. This approach contrasts with our previous work, which employed a combination of deterministic and stochastic pulses---generating random bits from a well-defined initial state~\cite{Rehm2023,Rehm2024,Dubovskiy2024,Sidi2025}. The write pulses are nanosecond in duration, placing the device operation near the ballistic spin-transfer regime, where the stochasticity arises primarily from thermal variations in the initial magnetization state~\cite{Bedau2010a,Liu2014}. In this regime, the switching probability depends on the product of pulse amplitude and duration---\emph{i.e.}, the total charge transmitted through they junction, a consequence of angular momentum conservation. We also note that operation at current amplitudes approximately twice the threshold current has been shown to minimize energy dissipation within the MTJ due to Joule heating~\cite{Bedau2010a}.

To analyze the device operation we consider the switching probabilities associated with pulses S1 and S2.
 The first pulse favors the ``1'' state with a switching probability $\Gamma_{01}$. Whereas, the second pulse favors the ``0'' state and has a switching probability of $\Gamma_{10}$. (Note that in the sMTJ case $\Gamma$ is a rate, \emph{e.g.}, flips/sec, whereas in the actuated stochastic device case $\Gamma$ is a switching probability.) Also, in analyzing the switching dynamics we assume that the pulses are independent events, that is that the layer magnetizations and device equilibrates between pulses. 
 
 We take $P_{\mathrm S 1}(N)$ to be the probability of being in the state ``1'' after the first pulse (S1) and $P_{\mathrm S 2}(N)$ is the probability of being in the state ``1'' after the second pulse (S2) in pulse cycle $N$. Then the following difference equations govern probabilities in the next cycle (see Eqs. 1 \& 2 in Ref.~\cite{Sun2024})
\begin{eqnarray}
    P_{\mathrm S 1}(N+1)&=&P_{\mathrm s 2}(N)\left[1-\Gamma_{10}\right], \nonumber 
\label{Eq:w1} \\
    P_{\mathrm S 2}(N)&=&P_{\mathrm S 1}(N)+\left[1-P_{\mathrm S 1}(N)\right]\Gamma_{01}. \nonumber  \label{Eq:w2} 
\end{eqnarray}
From these equations it is straightforward to show that in a steady state (after a large number of pulse sequences) the probability of readout of a ``1'' state is:
\begin{equation}
     p\equiv\lim_{N\rightarrow \infty} P_{\mathrm S 2}(N)=\frac{\Gamma_{01}}{\Gamma_{01}+\Gamma_{10}-\Gamma_{01}\Gamma_{10}}, \label{Eq:w3}
\end{equation}
independent of the initial pMTJ states, provided, of course, $0<\Gamma\leq 1$. We note that for $\Gamma_{01}=\Gamma_{10}\equiv \Gamma$, $p=1/(2-\Gamma)$, \emph{i.e.}, that $p \neq 1/2$. This is because we read the junction state after the second pulse.

\subsection{\label{sec:Experiment} Experimental Results}
\subsubsection{Quasistatic characteristics}
To test this concept we first measure the switching probabilities from known initial MTJ states to determine $\Gamma_{01}$ and $\Gamma_{10}$. We use pMTJs with nominal diameters of $40$ nm, and an parallel (P) and antiparallel (AP) resistances of  $2$ and $3.2\;\mathrm{k}\Omega$, respectively. The measurements were conducted at room temperature with no external magnetic applied field. The pMTJ effective energy barriers are $\Delta\approx26$ for AP$\rightarrow$P transition, and  $\Delta\approx51$ for P$\rightarrow$AP transition~\citep{Rehm2019}. 

We use an arbitrary waveform generator (Tektronix 7102) to produce the pulses sequence and a high-speed digitizer (Teledyne ADQ7DC) to probe the resulting state. The frequency of the input sequence is set to $104$ MHz. The pMTJ state is recorded and digitized during $10^8$ pulse periods using the Teledyne's FPGA~\cite{Sidi2025}. The switching probability $\Gamma$ is then computed as the number of changes of device states divided by the total number of events, \emph{i.e.}, $10^8$.

\begin{figure}
 \centering
 \includegraphics[width=0.45\textwidth]{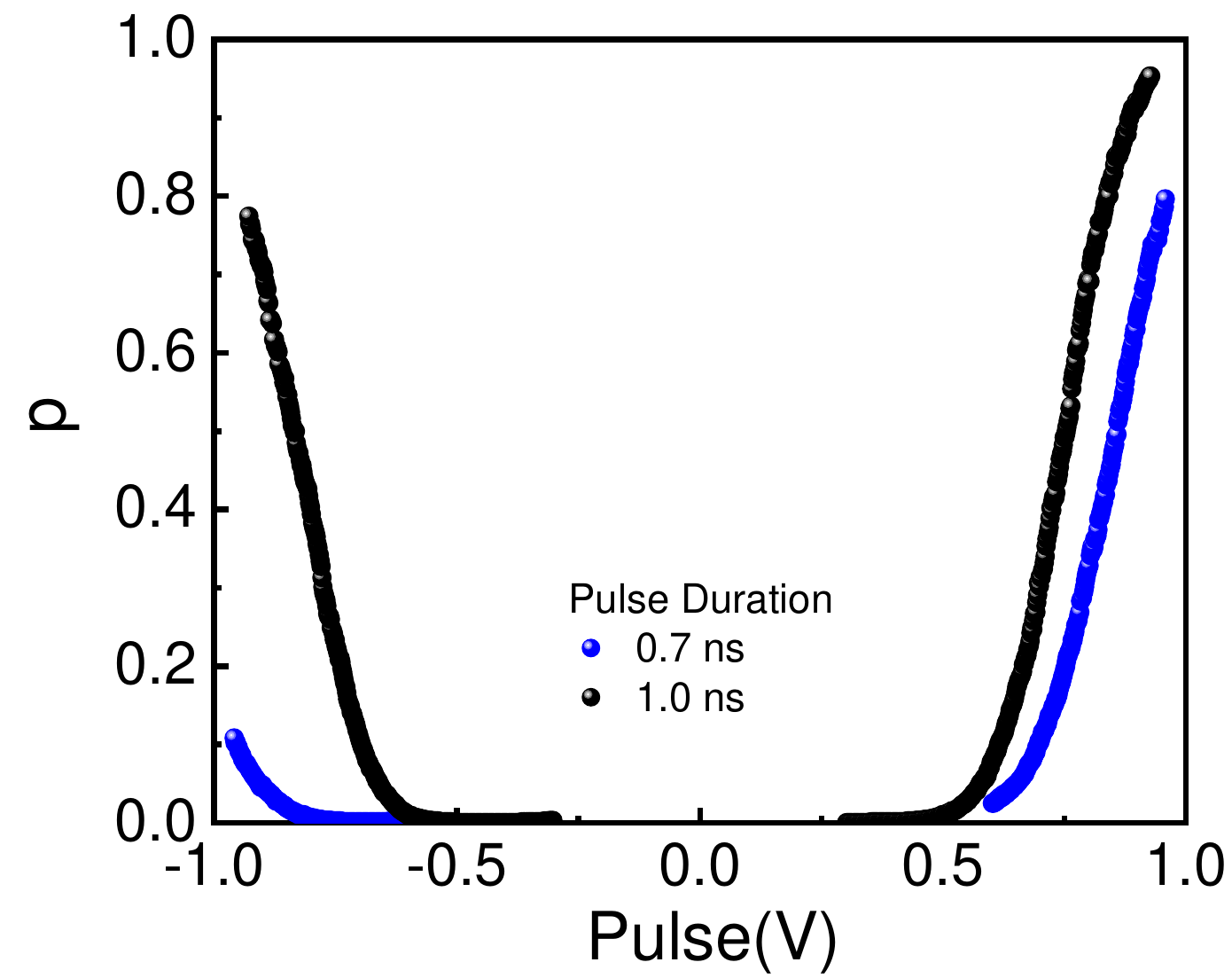}
 \caption{Measured switching probabilities. The switching probabilities $\Gamma_{01}$ (for positive voltage pulses) and $\Gamma_{10}$ (for negative voltage pulses) are plotted as a function of pulse amplitude for pulse durations of 0.7 and 1 ns. Each probability value is determined from $10^8$ switching attempts at the corresponding amplitude.}
 \label{Fig:Gammas}
\end{figure}
The switching probability dependence on voltage pulse starting from P and AP initial states are shown in Fig.~\ref{Fig:Gammas} for two pulse durations ($0.7$ and $1$ ns). These initial states were created using a deterministic first write pulse. In each case, the switching probability is, as expected, a monotonically increasing function of the pulse amplitude.
\begin{figure*}[t]
 \centering
 \includegraphics[width=1.0\textwidth]{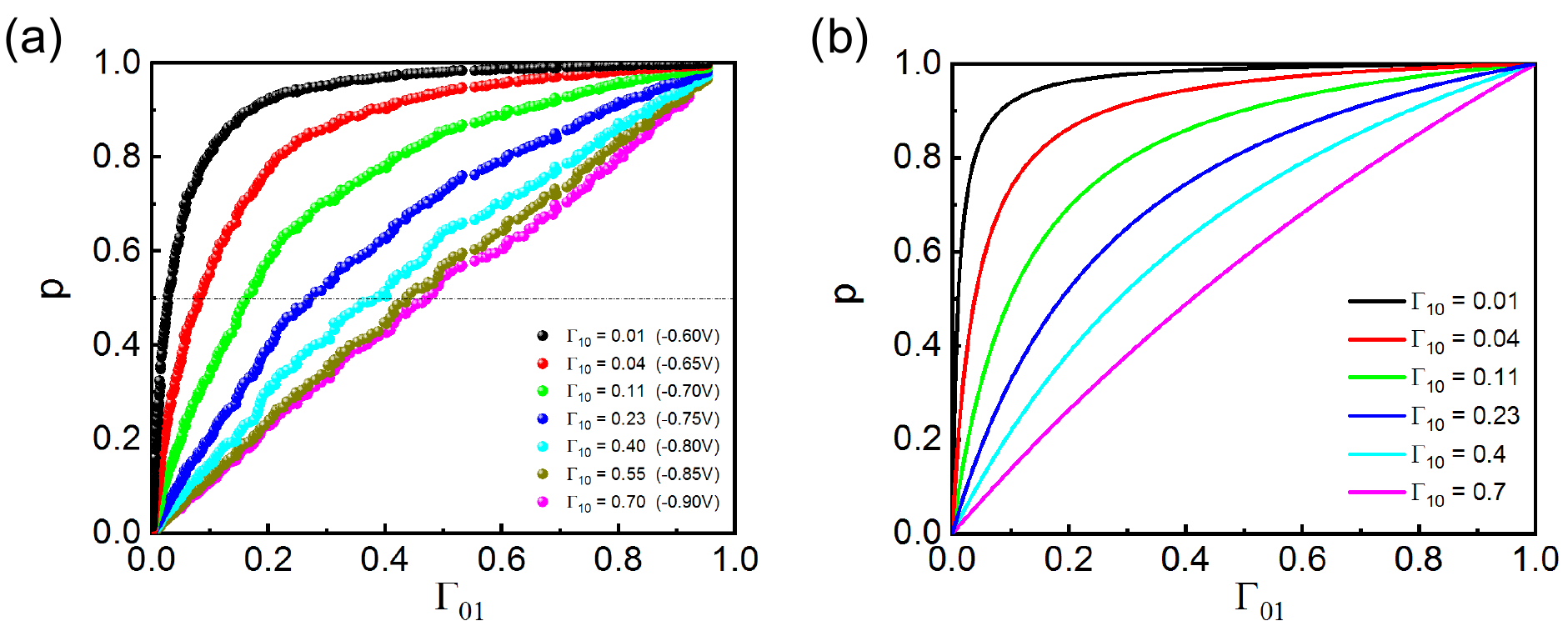}
\caption{(a) Measured output probability $p$ under AC pulse excitation with 1~ns stochastic pulses, plotted as a function of switching probability $\Gamma_{01}$ corresponding to the S1 pulse amplitude (inferred from the mapping in Fig.~\ref{Fig:Gammas}). The legend indicates the corresponding S2 switching probability $\Gamma_{10}$ with the corresponding pulse amplitudes in parentheses. The dashed line denotes $p = 0.5$, indicating the pulse combinations that achieve a balanced switching probability.
(b) Model predictions of $p$ based on Eq.~\ref{Eq:w2}, assuming independent switching events during each half-cycle.}
 \label{Fig:p}
\end{figure*} 

We now characterize the device response when both S1 and S2 pulses are stochastic by measuring the probability $p$ of reading a ``1'' (\emph{i.e.}, the AP state) as a function of the AC pulse amplitudes. The results in Fig.~\ref{Fig:p}(a) correspond to 1~ns duration stochastic pulses. In each cycle, two such pulses (S1 and S2) are separated by 2.6~ns, followed 0.4~ns later by a 4.6~ns read pulse. The x-axis is $\Gamma_{01}$, the S1 pulse amplitude mapped to the switching probability based on the device switching characteristics presented in Fig.~\ref{Fig:Gammas}.

The model prediction from Eq.~\ref{Eq:w2} shown in Fig.~\ref{Fig:p}(b) is in excellent agreement with the experimental data. When the switching probability $\Gamma_{10}$ is relatively high ($> 0.5$), the system exhibits frequent transitions between states and the probability $p$ is nearly linearly dependent on $\Gamma_{01}$. Conversely, at lower pulse amplitudes, both $\Gamma_{01}$ and $\Gamma_{10}$ become small, leading to increasingly nonlinear behavior. In this regime, the system undergoes less frequent stochastic oscillations between the $P$ and $AP$ states and small changes in switching probabilities, $\Gamma_{01}$ and $\Gamma_{10}$, lead to large changes in p. This case is discussed in more detail below.

\subsubsection{Temporal dynamics}
\begin{figure}
 \centering
 \includegraphics[width=0.5\textwidth]{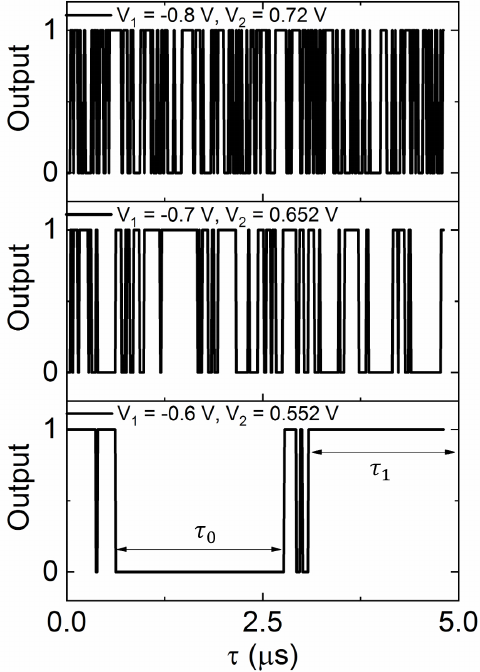}
 \caption{pMTJ digitized output under AC pulse excitation in real-time. The three vertical figures (a)-(c) show the variation in fluctuation frequency with the different amplitudes of the pulses $V_1$ and $V_2$. The probability of 1 or 0 in all three plots is $p = 0.5$. $\tau_0$, and $\tau_1$ refers to the dwell-times; the time spent in one state before transitioning to the other.} 
 \label{Fig:TRNG}
\end{figure}

We now focus on the temporal response of the device, where its most interesting and application-relevant behavior becomes clear. Figure~\ref{Fig:TRNG} presents measurements of the pMTJ state over a $5~\mu$s interval, sampled at 9.6 ns intervals. The device was driven by AC pulse excitations with S1 and S2 pulse amplitudes chosen to yield an average switching probability of $p = 0.5$, the pulse amplitude combinations indicated with the dashed line in Fig.~\ref{Fig:p}(a). Under these conditions, the traces reveal spontaneous fluctuations between the parallel (P, state = 0) and antiparallel (AP, state = 1) resistance states, highlighting the highly stochastic nature of the switching dynamics observed in A-sMTJs.

However, unlike conventional sMTJs---where the fluctuation rate is thermally driven and largely fixed---A-sMTJs offer full control over the fluctuation time scale through the amplitude (and, not shown here, the duration) of the driving pulses. Figure~\ref{Fig:TRNG}(a)–(c) illustrates that increasing the pulse amplitude results in substantially faster switching dynamics, demonstrating tunability of the fluctuation rate over several orders of magnitude.

\begin{figure*}
 \centering
 \includegraphics[width=1.0\textwidth]{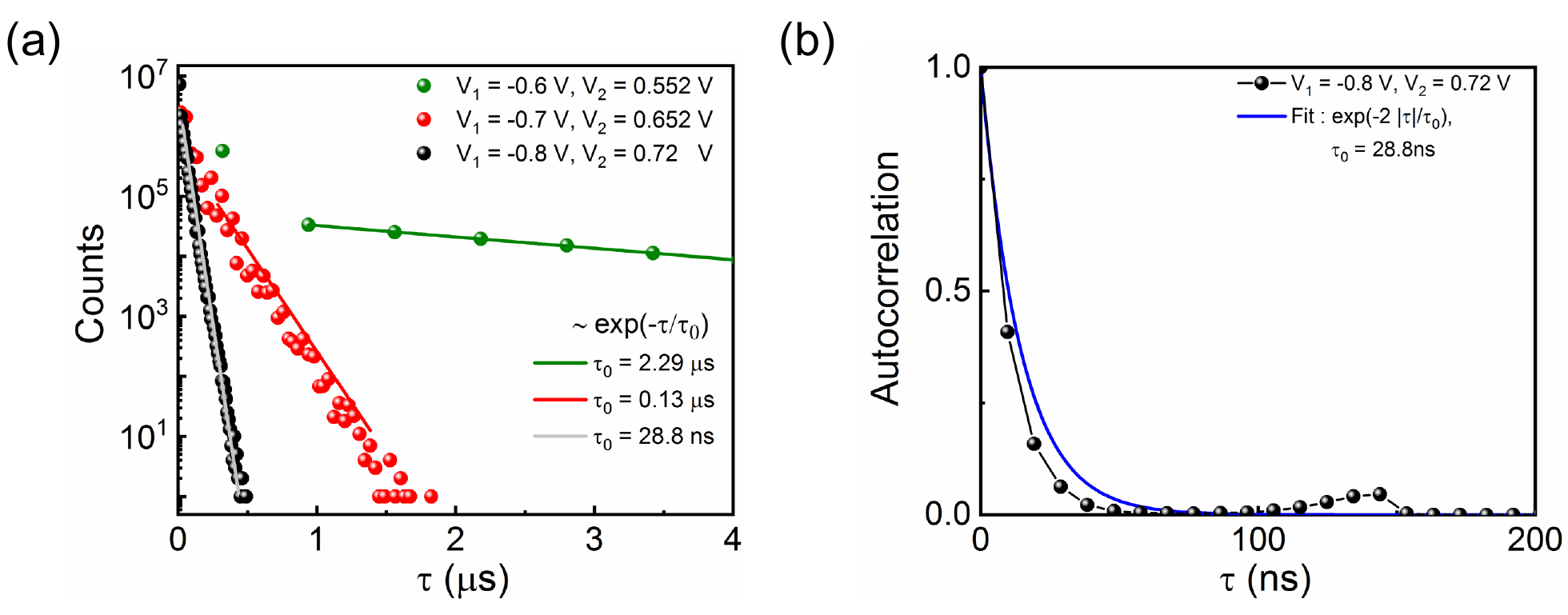}
 \caption{a) Histograms of $\tau_0$ dwell-times on a logarithm scale, and for different pulse amplitudes $V_1$ and $V_2$. The experimental data are the symbols and the lines are an exponential fit of the data. b) Normalized autocorrelation of the output for $V_1 = -0.8$ and $V_2 = 0.72$ V.}
 \label{Fig:distributions}
\end{figure*}
We characterize the device’s temporal behavior by measuring dwell times, calculated by converting the number of consecutive measurement cycles for which the readout remains the same into a duration. The dwell times, labeled $\tau_0$ and $\tau_1$, represent the durations the device remains in the ``0'' and ``1'' states, respectively, before switching occurs. Histograms of the measured dwell times are shown in Fig.~\ref{Fig:distributions}(a). As the pulse amplitudes increase, the occurrence of long dwell times (\emph{i.e.}, $\tau > 500~\mathrm{ns}$) is markedly reduced, reflecting the increase in fluctuation rate. Conversely, at lower pulse amplitudes—corresponding to slower fluctuation rates—the dwell times become significantly longer. The results show that $\tau_0$ can be tuned over more than two orders of magnitude, from approximately $2.3~\mu\mathrm{s}$ at $V_1 = -0.6~\mathrm{V}$ to $28.8~\mathrm{ns}$ at $V_1 = -0.8~\mathrm{V}$. 

The dwell-time histograms shown in Fig.~\ref{Fig:distributions}(a) for all three measurements exhibit an exponential decay, consistent with a distribution of the form $\propto \exp(-\tau/\tau_0)$. This behavior is consistent with a Poisson process characterized by a constant transition rate $1/\tau_0$.

To further test whether the experimental signal conforms to Poissonian statistics, we examined the autocorrelation of the A-sMTJ output and compared it to that of an ideal Poisson process, which follows the form $\exp(-2\tau/\tau_0)$. An example of this comparison is shown in Fig.~\ref{Fig:distributions}(b) for a signal recorded at $V_1 = -0.8\mathrm{V}$ over a 200 ns time window. The discrete autocorrelation was computed using the Python library \texttt{numpy}~\cite{Safranski2021} and normalized by its maximum value. The blue curve in Fig.~\ref{Fig:distributions}(b) shows the theoretical autocorrelation of a Poisson process with $\tau_0 = 28.8\mathrm{ns}$; the experimental data closely matches the exponential decay for time delays $\tau < 100~\mathrm{ns}$.

However, a secondary autocorrelation peak with weak amplitude appears at $\tau \approx 144~\mathrm{ns}$, deviating from the ideal Poisson prediction and suggesting temporal correlations between switching events. This secondary feature is consistently observed in dwell-time measurements under other experimental conditions, and has also been reported in conventional sMTJs~\cite{Soumah2025}. Its presence indicates that while the overall behavior is predominantly Poissonian, subtle memory effects or hidden correlations influence the switching dynamics. We are continuing to investigate the origin of these correlations.

\section{\label{sec:Discussion}Discussion}
These results establish a new class of tunable stochastic spintronic devices that combine thermal stability with externally controlled randomness, offering new capabilities for probabilistic computing, neuromorphic architectures, and true random number generation. In particular, we demonstratefd that pMTJs with thermally stable magnetic layers---originally developed and optimized for embedded spin-transfer torque magnetic random-access memory (STT-MRAM)---can be actuated to exhibit superparamagnetic-like behavior, producing highly tunable random telegraph noise under AC pulse excitation.

Compared to conventional superparamagnetic pMTJs, which rely on low energy barriers to enable thermally induced magnetization reversal, the actuated stochastic paramagnetic MTJ (A-sMTJ) offers key advantages in both stability and multifunctionality. While their higher energy barrier increase energy consumption, it also enhance resilience to external perturbations and preserve non-volatility. For instance, the stochastic dynamics of MTJ devices with higher barriers are less sensitive to variations in temperature, device geometry, and material properties, as discussed in Refs.~\cite{Rehm2024,Morshed2023}. Moreover, A-sMTJs enable active control over the device's operating mode—allowing it to function either as a deterministic switch or as a tunable stochastic element. In our implementation, the energy required per switching event is approximately $0.8$~pJ, compared to $\sim$2~fJ reported for thermally activated sMTJs~\cite{Vodenicarevic2017}. However, we anticipate that this energy cost can be substantially reduced by lowering the critical switching voltage---an optimization strategy that is  central to STT-MRAM development.

The overall results highlight the potential of standard pMTJs as versatile building blocks for unconventional computing architectures. Their multifunctionality offers a significant advantage, enabling the exploration of multiple computing paradigms—such as stochastic, deterministic, and in-memory operations—within a single chip architecture. Stable pMTJs have already been demonstrated as effective in-memory computing units, capable of performing analog multiply-and-accumulate (MAC) operations for tasks like image classification and recognition~\cite{Jung2022}. Moreover, the integration of in-memory computing with binary neural networks (BNNs) has been shown to reduce network complexity and improve memory density~\cite{Goodwill2022,Borders2024}. Further, networks of coupled stochastic MTJ have been shown to solve complex combinatorial optimization problems more efficiently than present methods~\cite{Chen2025}. Our demonstration of tunable stochastic behavior in standard pMTJs suggests that such devices could serve dual roles in future architectures—supporting both memory-centric and probabilistic or neuromorphic computation.

In summary, we have demonstrated that thermally stable pMTJs, commonly used in non-volatile memory technologies, can be used as tunable stochastic elements by applying nanosecond-scale alternating spin-transfer torque pulses. This actuated approach enables full control over the statistical properties of the output, including fluctuation rate and switching probability, while maintaining device stability. These features open the door to a new class of hybrid spintronic components that unify memory, logic, and randomness within a single physical framework. Future work will explore optimizing device parameters for energy efficiency, improving integration with CMOS circuits, and leveraging collective behavior in pMTJ arrays for large-scale probabilistic and neuromorphic systems.

\subsection*{Acknowledgements}
We acknowledge support from the Office of Naval Research (ONR) under Award
No. N00014-23-1-2771.  We thank Jonathan Z. Sun, Flaviano Morone and Dries Sels for helpful discussions of this work.

\end{document}